# Testing polynomial covariate effects in linear and generalized linear mixed models[*]

Mingyan Huang and Daowen Zhang

*Department of Statistics, North Carolina State University, Raleigh, NC 27695*
*e-mail:* mhuang@ncsu.edu
*e-mail:* dzhang2@stat.ncsu.edu

**Abstract:** An important feature of linear mixed models and generalized linear mixed models is that the conditional mean of the response given the random effects, after transformed by a link function, is linearly related to the fixed covariate effects and random effects. Therefore, it is of practical importance to test the adequacy of this assumption, particularly the assumption of linear covariate effects. In this paper, we review procedures that can be used for testing polynomial covariate effects in these popular models. Specifically, four types of hypothesis testing approaches are reviewed, i.e. R tests, likelihood ratio tests, score tests and residual-based tests. Derivation and performance of each testing procedure will be discussed, including a small simulation study for comparing the likelihood ratio tests with the score tests.

## Contents

---

**1. Introduction**

Linear mixed models (LMMs) [16] and their extension, generalized linear mixed models (GLMMs) [2; 28] are popular statistical models for analyzing correlated data, including longitudinal and clustered data often arising in biomedical research. An important feature of these models is that the conditional mean of the response given covariates and random effects, after transformed by a link function, is linearly related to the fixed covariate effects and random effects. The correctness of such model specification, especially the one on parametric linear covariate effects, has a significant impact on the validity of the subsequent statistical inference on the covariate effects. Therefore, it is of practical importance to check the adequacy of the assumption for the parametric linear covariate effects.

In order to evaluate the adequacy of a parametric covariate effect in a regression model, one common approach is to cast the problem in the hypothesis testing framework, where a broader class of models is selected as the alternatives. Nonparametric regression models, due to their flexibility and robustness in modeling the relationship between a response variable and explanatory variables, are often chosen as such alternatives. In practice, however, one rarely directly uses pure nonparametric regression models as alternatives because of the intrinsic infinite dimensional problem of nonparametric functions. To overcome such difficulties, various smoothing techniques, such as kernel smoothing and (penalized) spline smoothing, are often applied to estimate nonparametric functions, and the resulting estimates are then used as the alternatives for testing the adequacy of the parametric covariate effects. In doing so, the infinite dimensional alternatives are reduced to the ones with finite dimensions (or even one dimension in some special cases), which significantly simplifies the testing problems. For example, it is well-known that a nonparametric function estimated via penalized splines or smoothing splines has a mixed effects representation [3; 29; 30]. An appealing feature of using the mixed effects representation is that one can cast the hypothesis test of parametric against nonparametric covariate effects as a variance component test, which in most cases is a simple one-dimensional testing problem [30; 8]. The likelihood ratio and the score testing approaches reviewed here are mainly based on this mixed effects representation.

Alternatively, testing the adequacy of parametric covariate effects in LMMs and GLMMs can also be viewed as a goodness-of-fit problem. The residual based tests proposed by Pan and Lin [22] take this view. Specifically, these tests are "based on the cumulative sums of residuals over covariates or predicted values of the response variable" [22]. The major advantage of this approach is that it is valid against any alternatives that deviate from an assumed model.

For checking the adequacy of parametric covariate effects, we present here an overview on four types of hypothesis testing approaches that receive significant attention in the literature: R tests, likelihood ratio tests, score tests and residual-based tests. For each test, the derivation and performance are described first in the linear or generalized linear model framework, and then we mainly focus on their extensions to mixed models. The paper is organized as follows. Section 2 briefly introduces the models to be considered in this review. In Section 3, we review the four testing procedures. In Section 4, we present the results from a small simulation study to compare the performance of two popular testing procedures, the exact likelihood ratio test and the score test, based on mixed effects representation of (penalized) smoothing spline estimates of a nonparametric function. The paper is concluded in Section 5 with some discussion.

## 2. Generalized linear mixed models

In this section, we briefly introduce the models to be considered and notations to be used in this review. Since LMMs are special cases of GLMMs, we will only introduce GLMMs for longitudinal/clustered data. Suppose there are $m$ subjects (or clusters) in a data set. For the $i$th subject ($i = 1, 2, \ldots, m$), denote by $y_{ij}$ the $j$th measurement of the response variable ($j = 1, 2, \ldots, n_i$), and by $z_{ij}$, $s_{ij}$ and $t_{ij}$ the $j$th measurements of the $q$-dimensional covariates $z$, $p$-dimensional covariates $s$ (not including the intercept) and a scalar covariate $t$. Given subject-specific random effects $b_i$ and these covariate values, $y_{ij}$ is assumed to be independent and has a conditional density in an exponential family with conditional mean $\mu_{ij} = \mathrm{E}(y_{ij}|b_i)$ and conditional variance $\mathrm{var}(y_{ij}|b_i) = \omega_{ij}^{-1}\phi v(\mu_{ij})$, where $\omega_{ij}$ is a prior weight, $\phi$ is the dispersion parameter and $v(\cdot)$ is the variance function. The conditional mean $\mu_{ij}$ is assumed to be related to the covariates in the following GLMM [2]

$$g(\mu_{ij}) = s_{ij}^T \delta + m(t_{ij}, \gamma) + z_{ij}^T b_i, \tag{2.1}$$

where $g(\cdot)$ is a known monotone link function, $\delta$ are fixed effects of $s$, $m(t, \gamma) = \gamma_0 + \gamma_1 t + \cdots + \gamma_d t^d$ is the $d$-order ($d$ is a non-negative integer) polynomial covariate effect of $t$ with coefficients $\gamma_k$'s, and the random effects $b_i$ are usually assumed to have a multivariate normal distribution $\mathrm{N}\{0, D(\theta)\}$ with $\theta$ being the vector of unique parameters in the variance matrix of the random effects $b_i$.

Model (2.1) includes many popular models as special cases. When $g(\mu) = \mu$ and $y_{ij}$ is assumed to have a conditional normal distribution given random effects $b_i$, the model (2.1) reduces to an LMM considered by Laird and Ware [16]. Suppose we are confident about the parametric linear form $s_{ij}^T \delta$ in model (2.1) and are mainly concerned with the adequacy of $m(t, \gamma)$, the polynomial covariate effect of $t$. For this purpose, we consider the following semiparametric additive mixed models (SAMMs) proposed by Zhang and Lin [30] as alternative models to model (2.1)

$$g(\mu_{ij}) = s_{ij}^T \delta + f(t_{ij}) + z_{ij}^T b_i, \tag{2.2}$$

where $f(t)$ is a smooth but arbitrary function.

Denote $y = (y_{11}, \ldots, y_{1n_1}, \ldots, y_{m1}, \ldots, y_{mn_m})^T$, $S = (s_{11}, \ldots, s_{1n_1}, \ldots, s_{m1}, \ldots, s_{mn_m})^T$, $b = (b_1, \ldots, b_m)^T$, $Z_i = (z_{i1}^T, \ldots, z_{in_i}^T)^T$, $Z = \mathrm{diag}\{Z_1, \ldots, Z_m\}$, and $\mu = \mathrm{E}(y|b)$. In the next section, we discuss four procedures for checking the assumption that $f(t)$ is adequately represented by a polynomial function $m(t, \gamma)$.

## 3. Four testing procedures

### *3.1. R tests*

The R tests, discussed by Hastie and Tibshirani [15], were originally developed for testing smoothing parameters during the estimation of nonparametric functions through smoothing techniques for independent data. The idea of the R tests is analogous to the F statistic frequently used in linear regression models. One of the advantages of the R tests is their easy implementation, as under the null hypothesis the asymptotic distribution of the R statistic can be approximated by the chi-square distribution. However, the estimates of the degrees of freedom of chi-square distributions can be biased, and the resulting approximated critical values might be inaccurate. Moreover, the finite-sample distribution of the R statistic has not been studied [8].

A number of modifications on the original R tests have been made, including the correction of the bias of nonparametric estimates, reconstruction of the original test statistics and the corresponding distributions [1; 4; 8]. Here we briefly describe a version of R statistics proposed by Hardle et al. [13] under the generalized linear model (GLM) framework. They considered the following generalized partially linear model, a special case of SAMMs (2.2) for independent data ($n_i = 1$):

$$g(\mu_i) = s_i^T \delta + f(t_i). \tag{3.1}$$

Here, no random effect is required as $y_i$'s are independent, so the second subscript $j$ ($j = 1$) can be dropped for the simplicity of the notation.

Denote by $\tilde{\delta}$ and $\tilde{f}$ the estimates of $\delta$ and $f(t)$ under the null parametric model $H_0 : f(t) = m(t, \gamma)$, and by $\widehat{\delta}$ and $\widehat{f}$ the estimates under the alternative model $H_a : f(t) \neq m(t, \gamma)$. Let $\tilde{\mu}_i = g^{-1}\{s_i^T \tilde{\delta} + \tilde{f}(t_i)\}$ and $\widehat{\mu}_i = g^{-1}\{s_i^T \widehat{\delta} + \widehat{f}(t_i)\}$. The proposed R statistic for testing $H_0 : f(t) = m(t; \gamma)$ versus $H_a : f(t) \neq m(t; \gamma)$, is defined as

$$R = -2 \sum_{i=1}^{m} Q(\tilde{\mu}_i; \widehat{\mu}_i), \tag{3.2}$$

where $Q$ is the log quasi-likelihood function defined as $Q(\mu_i; y_i) = \int_{y_i}^{\mu_i} \frac{\omega_i(y_i - u)}{v(u)} du$. Note that here the non-parametric estimates are based on kernel smoothing methods instead of spline methods as discussed below. As Hardle et al. [13] pointed out, the usual likelihood ratio statistic $\mathcal{L}(\widehat{f}, \widehat{\delta}) - \mathcal{L}(\tilde{f}, \tilde{\delta})$, where $\mathcal{L}(f, \delta) = \sum_{i=1}^{m} Q(\mu_i; y_i)$, is not appropriate in this case as $\delta$ and $f(t)$ are estimated from two different likelihood functions. Under the null hypothesis, Hardle et al. [13] showed that the new R statistic has an asymptotic normal distribution, although such approximation typically does not work well. Hence Hardle et al. [13] proposed several sophisticated bootstrap-based approaches to obtain more accurate critical values for the R tests.

Sperlich and Lombardia [21] extended the above R statistic to test $H_0 : f(t) = m(t; \gamma)$ for a special SAMM with a random intercept only (*i.e.*, $z_{ij} = 1$).

The test statistic they proposed takes the following form:

$$R_{1w} = \sum_{i=1}^{m} \sum_{j=1}^{n_i} H\{\widehat{f}(t_{ij}), \widehat{\delta}\}\{\widehat{f}(t_{ij}) - \tilde{f}(t_{ij}) + s_{ij}^T(\widehat{\delta} - \tilde{\delta})\}^2 \pi(t_{ij}), \quad (3.3)$$

where $\pi(.)$ is a weight function which could be chosen empirically and

$$H\{f(t_{ij}), \delta\} = \frac{\partial}{\partial f} l(y_{ij}; f, \delta)^2,$$

with $l(y_{ij}; f, \delta) = \log f(y_{ij}|t, s, f, \delta)$, the log density of $y_{ij}$. The $R_{1w}$ statistic is based on "direct comparison" between estimates from nonparametric alternatives and estimates from null parametric models. Furthermore, Sperlich and Lombardia [21] showed that the theory of the asymptotic normal distribution from Hardle et al. [13] can be carried over to the test statistic $R_{1w}$. However, the asymptotic approximations often depart from the real finite sample distributions of the test statistics, which can lead to poor estimates of the critical values. Therefore, a number of bootstrap procedures were suggested to approximate the null distribution of the test statistic $R_{1w}$.

It can be immediately seen that construction of the R test statistic and its extension $R_{1w}$ for SAMMs involves the estimation of both the null and alternative models. Estimation of the null model may be relatively straightforward, however the model estimation under alternatives can be computationally intensive and sometimes challenging. The bootstrap procedure used to calculate the null distribution of the test statistics also requires significant computation time, which may limit the application scope of this testing approach.

### 3.2. Likelihood ratio tests

For testing a parametric versus nonparametric covariate effect, the likelihood ratio test (LRT) is a natural choice. The LRT has been popular in situations where we need to compare two nested models. However, extending the LRT to testing the adequacy of a parametric covariate effect is not straightforward. A considerable amount of work has been done in constructing likelihood ratio based test statistics for comparing parametric versus nonparametric covariate effects. Depending on how the nonparametric alternatives were specified and what types of smoothing techniques were used, a number of versions of likelihood ratio based testing procedures have been proposed. In this section, we review the LRTs based on the mixed model representation of a nonparametric function estimated using a (penalized) smoothing spline.

Crainiceanu and Ruppert [7] considered the exact LRT and restricted likelihood ratio test (RLRT) for testing whether the nonparametric function is a certain degree polynomial in the following partially linear model, which is a special case of SAMMs (2.2) and generalized partially linear models (3.1),

$$y_i = s_i^T \delta + f(t_i) + \epsilon_i, \quad (3.4)$$

where $\delta$ and $f(t)$ have the same definitions as before, $\epsilon_i$ are *i.i.d.* from $N(0, \sigma_\epsilon^2)$ and are assumed to be independent of $s_i$ and $t_i$. The nonparametric function $f(t)$ can be approximated through a penalized smoothing spline by the following spline function

$$f(t) = \gamma_0 + \gamma_1 t + \cdots + \gamma_d t^d + \sum_{k=1}^{K} a_k (t - \xi_k)_+^d, \qquad (3.5)$$

where $K$ is a non-negative integer, $\gamma = (\gamma_0, \cdots, \gamma_d)^T$, $a = (a_1, \cdots, a_K)^T$ are two sets of parameters, $(t)_+^d = t^d$ for $t > 0$ and zero otherwise, $\xi_1 < \cdots < \xi_K$ are fixed knots, and $\xi_k$ could be defined as the $k/(K+1)$th sample quantile of $t's$. In order for (3.5) to be a good approximation, $K$ is usually chosen to be large (such as 20), in which case it is not desirable to estimate $\gamma$ and $a$ directly. A penalized spline estimate of $f(t)$ is obtained by minimizing the following penalized least square equation

$$\sum_{i=1}^{m} \{y_i - f(t_i) - s_i^T \delta\}^2 + \frac{1}{\lambda} a^T \Sigma^{-1} a, \qquad (3.6)$$

where $\lambda$ is the smoothing parameter and $\Sigma$ is a pre-specified roughness penalty matrix, usually taken to be the identity matrix $\Sigma = I_{K \times K}$.

Let $A$ be the $m \times (d+1)$ matrix with the $i$th row $A_i = (1, t_i, \cdots, t_i^d)$ and $B$ be the $m \times K$ matrix with the $i$th row $B_i = [(t_i - \xi_1)_+^d, \cdots, (t_i - \xi_K)_+^d]$. The penalized least square equation (3.6) suggests that $f(t)$ has a mixed effects representation $f = A\gamma + Ba$, where $f = \{f(t_1), f(t_2), \ldots, f(t_m)\}^T$, $\gamma$ is considered as fixed effects and $a$ is regarded as random effects having the distribution $a \sim N(0, \sigma_a^2)$ with $\sigma_a^2 = \lambda \sigma_\epsilon^2$. Denote $\beta = (\delta^T, \gamma^T)^T$ and $X = [S|A]$ where $S$ is the $m \times p$ matrix with the $i$th row $s_i^T$. Then the original partially linear model has the equivalent linear mixed model representation

$$Y = X\beta + Ba + \epsilon. \qquad (3.7)$$

It can be clearly seen from the penalized spline expression (3.5) that generally $f(t)$ is a polynomial of degree $d-h$ ($h = 0, 1, \ldots, d$) if $\gamma_{d-h+1} = \cdots = \gamma_d = 0$ and $a_1 = \cdots = a_K = 0$, which is equivalent to $\gamma_{d-h+1} = \cdots = \gamma_d = 0$ and $\sigma_a^2 = 0$ (or $\lambda = 0$) using the linear mixed model representation. Therefore, testing whether the covariate effect of $t$ is a $(d-h)$-degree polynomial is equivalent to testing $H_0 : \gamma_{d-h+1} = \cdots = \gamma_d = 0, \sigma_a^2 = 0$ ($\lambda = 0$) versus $H_a : \gamma_{d-h+1} \neq 0$ or $\cdots$ or $\gamma_d \neq 0$ or $\sigma_a^2 > 0$ ($\lambda > 0$) if the mixed model representation of a penalized smoothing spline is used. One approach proposed by Crainiceanu and Ruppert [7] for testing this hypothesis is the LRT using the log-likelihood of $\beta, \sigma_a^2$ and $\sigma_\epsilon^2$ from the mixed model representation (3.7)

$$\ell(\beta, \sigma_a^2, \sigma_\epsilon^2; Y) = -\frac{1}{2} \log |V| - \frac{1}{2} (Y - X\beta)^T V^{-1} (Y - X\beta),$$

where $V = \sigma_a^2 BB^T + \sigma_\epsilon^2 I_{m \times m}$ is the marginal variance of $Y$ under the model (3.7). In the case where $h = 0$, the testing problem becomes a variance component test, i.e. $H_0 : \sigma_a^2 = 0$ versus $H_a : \sigma_a^2 > 0$. Besides the LRT, an alternative choice for testing this particular hypothesis is to use the following REML function

$$\ell_R(\sigma_a^2, \sigma_\epsilon^2; Y) = -\frac{1}{2}\log|V| - \frac{1}{2}\log|X^TV^{-1}X| - \frac{1}{2}(Y - X\widehat{\beta})^TV^{-1}(Y - X\widehat{\beta}),$$

where $\widehat{\beta} = (X^TV^{-1}X)^{-1}X^TV^{-1}Y$. This method is abbreviated by RLRT.

As pointed out by Crainiceanu and Ruppert [7], under $H_0$ the LRT or RLRT asymptotically does not follow a $0.5\chi_0^2 + 0.5\chi_1^2$ mixture chi-square distribution as suggested by Self and Liang [23] and Stram and Lee [25]. Instead, the LRT or RLRT asymptotically follows a mixture of $\chi_0^2$ and $\chi_1^2$ with a much heavier mass on $\chi_0^2$. A simple and fast algorithm was also proposed to sample the exact null distribution of the LRT or RLRT, which is summarized as follows [7]:

Step 1: Generate a grid of $\lambda$ values where $0 = \lambda_1 < \lambda_2 < \cdots < \lambda_n$.
Step 2: Simulate $K$ independent random variables $w_1^2, \cdots, w_K^2$ from the $\chi_1^2$. Let $S_K = \sum_{s=1}^K w_s^2$.
Step 3: Independently simulate $X_{m,K,d} = \sum_{s=K+1}^{m-p-d-1} w_s^2$ with $w_s^2 \sim \chi_1^2$.
Step 4: When $h \neq 0$, independently simulate $X_h = \sum_{s=1}^h u_s^2$ with $u_s^2 \sim \chi_1^2$.
Step 5: For every grid point $\lambda_i$ calculate

$$N_m(\lambda_i) = \sum_{s=1}^K \frac{\lambda_i \mu_{s,m}}{1 + \lambda_i \mu_{s,m}} w_s^2$$

$$D_m(\lambda_i) = \sum_{s=1}^K \frac{w_s^2}{1 + \lambda_i \mu_{s,m}} + X_{m,K,d}.$$

Step 6: Obtain $\lambda_{max}$ that maximizes $f_m(\lambda_i)$ over $\lambda_1, \cdots, \lambda_n$, where

$$f_m(\lambda) = m\log\left\{1 + \frac{N_m(\lambda)}{D_m(\lambda)}\right\} - \sum_{s=1}^K log(1 + \lambda\zeta_{s,m}).$$

Step 7: Compute the LRT statistic $LRT_m = f_m(\lambda_{max}) + m\log(1 + \frac{X_h}{S_K + X_{m,K,d}})$, or $LRT_m = f_m(\lambda_{max})$ if $h = 0$. For the case of RLRT, compute

$$RLRT_m = \sup_{\lambda \geq 0}\left[(m - p - d - 1)\log\left\{1 + \frac{N_m(\lambda)}{D_m(\lambda)}\right\} - \sum_{s=1}^K \log(1 + \lambda\mu_{s,m})\right].$$

Step 8: Repeat steps 2–7.

Here $\mu_{s,m}$ and $\zeta_{s,m}$ are defined to be the $K$ eigenvalues of the $K \times K$ matrices $Z^TP_0Z$ and $Z^TZ$ respectively, where $P_0 = I_m - X(X^TX)^{-1}X^T$.

In a recent (unpublished) paper, Claeskens et al. [5] adapted the idea of Crainiceanu and Ruppert [7] and explored the advantages of wavelets for estimating nonparametric smooth functions over the use of penalized splines in partially linear models for independent data. Two asymptotic distribution theorems were developed for the test statistics proposed therein, and simulation results showed that the wavelet-based test has better performance than the penalized spline based test in some situations. They also extended the wavelet based test to the cases of simultaneously testing several polynomial covariate effects.

For testing generalized linear models with a single covariate $t$ for independent discrete data, Liu et al. [20] proposed three methods which are "based on the connection between smoothing spline models and Bayesian models", assuming $f(t)$ in model (3.1) to have the following Bayesian expression

$$f(t) = \gamma_0 + \gamma_1 t + \cdots + \gamma_d t^d + \tau^{1/2} W(t),$$

where $\gamma_0, \gamma_1, \ldots, \gamma_d$ have flat prior, and $W(t)$ is the $d$-order Wiener process. Under this Bayesian model, they extended the generalized maximum likelihood ratio (GML) test of Wahba [27] to test the adequacy of a generalized linear model, which is equivalent to $H_0 : \tau = 0$. The test statistic of the GML test proposed by Liu et al. [20] is constructed as

$$t_{GML} = \frac{sup_\phi L(0, \phi|y)}{sup_{\tau,\phi} L(\tau, \phi|y)}, \qquad (3.8)$$

where $L(\tau, \phi|y)$ denotes the marginal density of $y$ under this Bayesian model. Obviously, under the mixed model representation of a smoothing spline estimate of a nonparametric function $t_{GML}$ is essentially a LRT.

One difficulty with the GML test is that there is no closed form expression for $L(\tau, \phi|y)$, and the test statistic can only be approximated numerically [20]. Secondly, it is nearly impossible to analytically derive the null distribution of the test statistic as its distribution depends on some unknown parameters. To overcome this difficulty, Liu et al. [20] suggested two approaches to approximating the exact null distribution of the test statistic. One is the usual bootstrap procedure which is computationally intensive. The other approach is the so called empirical approximation method, which was considered superior to the bootstrap-based method.

It should be noted that the testing procedures based on the likelihood ratio are all proposed for models for independent data. Although conceptually they can be extended to SAMMs for longitudinal/clustered data, there are at least two major obstacles. First the calculation of the likelihood is even more complicated under the alternative using the mixed model representation of a (penalized) smoothing spline estimate of a nonparametric function. Secondly, it may not be easy to extend the algorithm of Crainiceanu and Ruppert [7], originally proposed for simulating the exact distribution of the LRT in a partially linear model, to SAMMs or even LMMs for longitudinal/clustered data. More future research is needed in this area.

### *3.3. Score tests*

In generalized linear models, score tests have been used for testing the overdispersion and heterogeneity of outcomes [10; 24]. Lin [19] extended score tests to

GLMMs, in which a global score test as well as individual score tests were proposed to test the null hypotheses of all zero random-effect variance components and individual zero random-effect variance components respectively.

Zhang and Lin [30] considered the problem of testing the nonparametric function $f(t)$ in model (2.2) being a $d$-order polynomial. They first estimated $f(t)$ by a $d$-order smoothing spline and expressed $f$ with a mixed effects representation, similar to the one in Section 3.2 for a penalized smoothing spline

$$f = T\gamma + \Sigma a, \qquad (3.9)$$

where $f = f(t^0)$, $t^0$ is the vector formed by distinct $\{t_{ij}\}$'s, $T$ is a matrix formed by zero to the $d$th polynomials of $t^0$ with corresponding coefficients $\gamma$, $\Sigma$ is a smoothing matrix, and $a \sim N(0, \tau I)$. Note that this mixed effects representation is basically the same as the Bayesian expression presented in Section 3.2.

Denote by $N$ the incidence matrix mapping $t^0$ to $\{t_{ij}\}$'s, and define $X = (NT, S)$, $B = N\Sigma$. Then under the mixed effects representation (3.9), SAMM (2.2) becomes the following GLMM

$$g(\mu) = X\beta + Ba + Zb, \qquad (3.10)$$

where $\beta = (\gamma^T, \delta^T)^T$ are the new fixed effects and $(a, b)$ are the new random effects.

As described in the earlier sections, testing $f(t)$ in SAMM (2.2) being a $d$-order polynomial is equivalent to testing $H_0 : \tau = 0$ in the induced GLMM (3.10). Zhang and Lin [30] adapted the idea of Lin's [19] variance component score tests to test $H_0 : \tau = 0$. However, they pointed out that the score tests proposed by Lin [19] for testing zero variance components in GLMMs cannot be used directly for testing $H_0 : \tau = 0$. They proposed a scaled chi-squared approximation to the test statistic.

Denote by $\psi = (\theta^T, \phi)$ the nuisance parameter vector, and by $\ell_M(\tau, \psi)$ the marginal log-likelihood function of $\tau$ and $\psi$ (by integrating out random effects $a$, $b$ and fixed effects $\beta$). Then under the induced GLMM (3.10), the score $\mathcal{U}_\tau$ for testing $H_0 : \tau = 0$ takes the following form

$$\begin{aligned} \mathcal{U}_\tau(\widehat{\psi}) &= \left.\frac{\partial \ell_M(\tau, \psi; y)}{\partial \tau}\right|_{\tau=0,\widehat{\psi}} \qquad (3.11) \\ &\approx \left.\frac{1}{2}\{(Y - X\beta)^T V^{-1} N\Sigma N^T V^{-1}(Y - X\beta) - tr(PN\Sigma N^T)\}\right|_{\widehat{\beta},\widehat{\psi}}, \end{aligned}$$

where $\widehat{\beta}$ is the MLE of $\beta$ and $\widehat{\psi}$ is the REML-type of estimate of $\psi$ under the following null GLMM (3.12), and $Y = X\beta + Zb + \Delta(y - \mu)$ is the working vector from the null GLMM

$$g(\mu) = X\beta + Zb, \qquad (3.12)$$

where $P = V^{-1} - V^{-1}X(X^T V^{-1}X)^{-1}X^T$, $V = W^{-1} + ZGZ^T$, $G = \text{diag}\{D, \ldots, D\}$, $\Delta = \text{diag}\{g'(\mu_{ij})\}$, $W = \text{diag}\{w_{ij}\}$ and $w_{ij} = \{\phi\omega_{ij}^{-1}v(\mu_{ij})[g'(\mu_{ij})]^2\}^{-1}$. Note that model (3.12) is the matrix representation of the original GLMM (2.1).

Because of the special structure of $\Sigma$, Zhang and Lin [30] found that the score $\mathcal{U}_\tau(\widehat{\psi})$ does not follow an asymptotic normal distribution. Write $\mathcal{U}_\tau(\psi)$ as $\mathcal{U}_\tau(\psi) = U_\tau(y;\psi) - e(\psi)$, where $U_\tau(y;\psi)$ and $e(\psi)$ denote the first and the second terms of the above score, and define $\psi_0$ as the true value of $\psi$ under $H_0 : \tau = 0$. Zhang and Lin [30] showed that the null distribution of $U_\tau(y;\psi_0)$ is approximately equal to the one of weighted chi-squared random variables and can be well approximated by a scaled chi-squared distribution. Since the expectation of $\mathcal{U}_\tau(\psi)$ is an increasing function of $\tau$, larger values of $\mathcal{U}_\tau(\widehat{\psi})$ give more evidence against $H_0$, which indicates that the score test should be one-sided.

Compared with the LRTs, one major advantage of using the score test statistic $U_\tau(y;\widehat{\psi})$ is its easy implementation, as it can be calculated directly by fitting a GLMM (under the null hypothesis) rather than a SAMM. In addition, the critical values can be directly approximated from the regular chi-square distribution. Therefore, it is not necessary to derive the distribution of the test statistics under the null hypothesis as often required by the LRTs. Secondly, as SAMMs encompass a broad class of statistical models, the above score test can be applied in many situations, such as independent Gaussian data [6], clustered Gaussian or binary data, etc. For clustered data, the implementation of the LRTs can be very difficult as expensive computation is needed to approximate the null distribution of the test statistics.

The simulation results showed that the score test statistic above performs very well for Gaussian outcomes, less so for binary data due to the poor approximation of the Laplace method in calculating the score statistic, but improves rapidly as the binomial denominator increases [30].

### 3.4. Residual based tests

Inspired by the idea of residual plots for checking the goodness-of-fit of regression models, recently Pan and Lin [22] introduced a graphical and numerical approach to assess the adequacy of GLMMs. These methods are "based on the cumulative sums of residuals over covariates or predicted values of the response variable" [22] and are the further extensions of the work by Su and Wei [26] and Lin et al. [18].

Denote by $\mu_{ij}(\beta,\theta,\phi) = \mathrm{E}(y_{ij})$, the marginal mean of $y_{ij}$ and define residual $e_{ij}$ as $e_{ij} = y_{ij} - \widehat{\mu}_{ij}$, where $\widehat{\mu}_{ij} = \mu_{ij}(\widehat{\beta},\widehat{\theta},\widehat{\phi})$, and $\widehat{\beta},\widehat{\theta},\widehat{\phi}$ are the estimates of the corresponding parameters under the original GLMM (2.1) or model (3.12) in the matrix notation. Pan and Lin [22] then considered the following two classes of stochastic processes

$$W(x) = m^{-1/2} \sum_{i=1}^{m} \sum_{j=1}^{n_i} I(x_{ij} \leq x) e_{ij},$$

$$W_g(r) = m^{-1/2} \sum_{i=1}^{m} \sum_{j=1}^{n_i} I(\widehat{\mu}_{ij} \leq r) e_{ij},$$

where $x = (x_1, \cdots, x_p)^T$, $r \in \mathcal{R}$, $I(x_{ij} \leq x) = I(x_{1ij} \leq x_1, \cdots, x_{pij} \leq x_p)$, and $x_{kij}$ is the $k$th component of $x_{ij}$.

Under the assumed GLMM, these stochastic processes converge in distribution to zero-mean Gaussian processes, which can be simulated through Monte Carlo techniques. Each observed cumulative-sum process $W(x)$ or $W_g(r)$ can then be compared, both visually and analytically, to a certain zero-mean Gaussian process. If the assumed GLMM is a reasonable model for the given data, the cumulative-sum processes would behave like white noise. Therefore, any abnormal departure of $W(x)$ or $W_g(r)$ from the zero-mean Gaussian processes would be an indication of model mis-specification. The main advantage of this testing approach is that there is no need to specify the alternatives, therefore it can be used to test whether or not $f(t)$ in SAMM (2.2) can be adequately represented by a polynomial function. Nevertheless this test may be less powerful compared to the other procedures specifically designed for testing $f(t)$.

Introduced by Fan and Huang [11], another residual based test is the so called "adaptive Neyman test". Although the test statistic is constructed in a completely different way, the basic idea is similar to the one described above, i.e. if a parametric model fits data well, the residuals should fluctuate around 0. They focused on the classical nonparametric model, which is $y = f(x) + \epsilon$ with $\epsilon \sim N(0, \sigma^2)$. Under the null hypothesis $f(\cdot) = m(\cdot, \gamma)$ for some $\gamma$, where $m(\cdot, \gamma)$ belongs to a given parametric family, the resulting residuals are given as $\widehat{\epsilon}_i = y_i - m(x_i, \widehat{\gamma})$, $i = 1, \cdots, n$, where $\widehat{\gamma}$ is the estimate of $\gamma$ under the assumed model. Denote $\widehat{\epsilon} = (\widehat{\epsilon}_1, \ldots, \widehat{\epsilon}_n)$, then $\widehat{\epsilon}$ is nearly independently and normally distributed with mean vector $\eta = (\eta_1, \cdots, \eta_n)^T$, where $\eta_i = f(x_i) - m(x_i, \gamma_0)$ and $\gamma_0$ is the convergent limit of $\widehat{\gamma}$. Thus, the testing problem can be constructed as $H_0 : \eta = 0$ versus $H_a : \eta \neq 0$. Fan and Huang [11] adopted the adaptive Neyman test to this testing problem. The adaptive Neyman test statistic is constructed based on the Fourier transform of the residuals $\widehat{\epsilon}$ with its exact null distribution being generated through simulations.

As mentioned earlier, the adaptive Neyman test has only been studied in partially linear models. So, extending it to LMMs or GLMMs could potentially be a future research direction.

## 4. Comparison between the exact likelihood ratio and the score tests

In this paper, we provided an overview of the four types of testing approaches. Among them, likelihood ratio and score tests have been widely used in a variety of hypothesis testing problems. To our knowledge, however, no comparison between these two tests has been investigated for the current situation, i.e. testing a parametric covariate effect against a nonparametric covariate effect. Here, we conduct a small simulation study to evaluate and compare the performance of these two popular testing procedures. For illustration purposes, we consider testing the linearity of covariate effects under the partially linear model framework, i.e. whether $f(t)$ is a linear function of $t$ in model (3.4). Following the penalized spline, we formulate the exact LRT (named as LRT1), RLRT and the score test as variance component tests based on the mixed model representation (3.7) as discussed above. In additon, for testing the same null hypothesis, we also formulate the exact LRT in a different way (named as LRT2) by modeling the alternative through a quadratic spline. In the latter case, we are testing whether $f(t)$ is a $(d - h)$-degree polynomial of $t$ with $d = 2$ and $h = 1$.

Since no exact LRT or RLRT has been developed for mixed models for longitudinal/clustered data, we only consider partially linear models for independent data even though Zhang and Lin's [30] procedure is applicable to more complicated models.

Data in this simulation are generated from the following partially linear model

$$y_i = s_{i1}\beta_1 + s_{i2}\beta_2 + f(t_i) + \epsilon_i, \quad i = 1, 2, \ldots, m$$

where $s_{i1}$ is generated from N$(0, 0.3)$, $s_{i2}$ is generated from N$(0, 0.4)$, $t_i$'s are equally spaced distinct points in [0,1], and $\epsilon_i \sim$ N$(0, \sigma^2)$. The true values of $\beta_1$ and $\beta_2$ are set to be 1.3 and 0.45 respectively. The values of $\sigma$ are 0.25 and 0.5, and the sample size m is taken to be 50 and 100. A total of five different functions of $f(t)$ are considered, i.e., $f_c(t) = (0.25c)t \cdot exp(2 - 2t) - t + 0.5$, for $c = (0, 1, 2, 3, 4)$ [30]. Note that when $c = 0$, $f_c(t)$ is a linear function of $t$ and $f_c(t)$ deviates further from linearity with increasing $c$. We apply the exact LRT1, LRT2, RLRT and the score testing procedures to each simulated data set. The simulation results are based on 1000 Monte Carlo simulation runs.

For testing the null hypothesis that $f(t)$ is a linear function of $t$, the size and power of each testing procedure are calculated by setting $c = 0$ and $c \neq 0$ respectively. When a penalized spline is used to estimate $f(t)$ as in the LRT or RLRT, the number of knots for the penalized spline is set to be 20. For the score testing procedure, the smoothing matrix $\Sigma$ is from a natural smoothing spline.

The simulation results are presented in the Table 1 ($m = 50$) and Table 2 ($m = 100$), where the nominal levels are set to be 0.05 and 0.1. Regarding the empirical size, our simulation results show that the exact LRT2, RLRT and the score test are all close to the nominal levels. The empirical size of the LRT1, however, stays unchanged even if the nominal level increases from 0.05 to 0.1. Overall the increased sample size brings the empirical sizes of all these tests closer to the nominal levels, whereas the error noise seems to have not much influence on them. With respect to the power, all tests show decreased power as the error variance increases. As expected, the increased sample size improves the overall power. Note that the powers of the LRT1 are also unchanged as the nominal level increases, which implies that the simulated critical values for the LRT1 may not be accurate with a moderate number of Monte Carlo simulation runs. In general, our simulation indicates that the LRT2, RLRT and score test are more powerful than the LRT1, with the score test slightly out-performing the exact LRT2 and RLRT.

In comparing to likelihood ratio based tests, the score test has at least two main advantages. First the exact LRT (LRT1 and LRT2) and RLRT are computationally much more intensive than the score test, as deriving the null distributions of the LRT and RLRT statistics requires simulation in each run. The computing time of the exact LRT and RLRT in this simulation is 50 times more than that of the score test. Secondly, the exact LRT and RLRT have not yet been developed for more complicated models such as LMMs and GLMMs, whereas the score testing procedure is flexible and can be adapted to many modeling situations. For simplicity, only the linearity test is considered in the current simulation; however in practice, one might be interested in testing higher-order polynomial covariate effects (*i.e.* $d > 1$), which can be easily carried out by using a different $d$. Overall we consider the score test is a better choice than the LRT and RLRT.

TABLE 1
*Empirical sizes and powers of the four tests in testing the linearity of covariate effects in model (3.4) where $m = 50$*

| nominal | $\sigma$ | Test | Size | Power | | | |
|---|---|---|---|---|---|---|---|
| level | | | $c=0$ | $c=1$ | $c=2$ | $c=3$ | $c=4$ |
| 0.05 | 0.25 | LRT1 | 0.032 | 0.152 | 0.696 | 0.991 | 1.000 |
| | | LRT2 | 0.049 | 0.419 | 0.935 | 0.999 | 1.000 |
| | | RLRT | 0.067 | 0.419 | 0.927 | 1.000 | 1.000 |
| | | Score | 0.066 | 0.443 | 0.948 | 1.000 | 1.000 |
| | 0.5 | LRT1 | 0.066 | 0.094 | 0.224 | 0.473 | 0.782 |
| | | LRT2 | 0.047 | 0.135 | 0.412 | 0.737 | 0.923 |
| | | RLRT | 0.050 | 0.123 | 0.404 | 0.720 | 0.915 |
| | | Score | 0.060 | 0.158 | 0.448 | 0.762 | 0.936 |
| 0.1 | 0.25 | LRT1 | 0.032 | 0.152 | 0.696 | 0.991 | 1.000 |
| | | LRT2 | 0.115 | 0.548 | 0.962 | 0.999 | 1.000 |
| | | RLRT | 0.138 | 0.545 | 0.970 | 0.999 | 1.000 |
| | | Score | 0.124 | 0.560 | 0.972 | 1.000 | 1.000 |
| | 0.5 | LRT1 | 0.066 | 0.094 | 0.224 | 0.473 | 0.782 |
| | | LRT2 | 0.093 | 0.230 | 0.545 | 0.838 | 0.961 |
| | | RLRT | 0.103 | 0.213 | 0.531 | 0.832 | 0.960 |
| | | Score | 0.104 | 0.242 | 0.565 | 0.859 | 0.970 |

TABLE 2
*Empirical sizes and powers of the four tests in testing the linearity of covariate effects in model (3.4) where $m = 100$*

| nominal | $\sigma$ | Test | Size | Power | | | |
|---|---|---|---|---|---|---|---|
| level | | | $c=0$ | $c=1$ | $c=2$ | $c=3$ | $c=4$ |
| 0.05 | 0.25 | LRT1 | 0.044 | 0.217 | 0.950 | 1.000 | 1.000 |
| | | LRT2 | 0.053 | 0.675 | 0.994 | 1.000 | 1.000 |
| | | RLRT | 0.052 | 0.661 | 0.995 | 1.000 | 1.000 |
| | | Score | 0.052 | 0.691 | 0.997 | 1.000 | 1.000 |
| | 0.5 | LRT1 | 0.068 | 0.115 | 0.364 | 0.810 | 0.988 |
| | | LRT2 | 0.059 | 0.240 | 0.681 | 0.956 | 0.999 |
| | | RLRT | 0.054 | 0.221 | 0.670 | 0.959 | 0.999 |
| | | Score | 0.062 | 0.249 | 0.697 | 0.963 | 0.999 |
| 0.1 | 0.25 | LRT1 | 0.044 | 0.217 | 0.950 | 1.000 | 1.000 |
| | | LRT2 | 0.109 | 0.778 | 0.998 | 1.000 | 1.000 |
| | | RLRT | 0.102 | 0.762 | 0.999 | 1.000 | 1.000 |
| | | Score | 0.107 | 0.779 | 1.000 | 1.000 | 1.000 |
| | 0.5 | LRT1 | 0.068 | 0.115 | 0.364 | 0.810 | 0.988 |
| | | LRT2 | 0.103 | 0.353 | 0.781 | 0.975 | 1.000 |
| | | RLRT | 0.112 | 0.336 | 0.777 | 0.982 | 1.000 |
| | | Score | 0.111 | 0.363 | 0.798 | 0.983 | 1.000 |

## 5. Summary

We overview the main development of the four types of testing approaches used for testing a parametric covariate effect versus a nonparametric covariate effect. A considerable amount of work has been done with the LRTs under linear or generalized linear models. The likelihood based tests perform very well for independent data in finite sample situations. However, these test statistics can be difficult to compute in a more complex model, as both the parametric and nonparametric models need to be estimated.

In addition, deriving the null distributions of those test statistics can be challenging. Therefore, it is not straightforward to extend the existing LRTs or RLRTs to LMMs and GLMMs. Compared to the LRTs or RLRTs, the score statistics are easy to compute, usually show good performance and are applicable to both LMMs and GLMMs. Further study may be needed to investigate the properties of the score tests for small samples. The R tests are likelihood-ratio-based tests, hence they share the same advantages and disadvantages as the LRTs. The recently developed residual-based test [22] can be considered as an omnibus test for detecting model mis-specification and can be used to test the adequacy of a polynomial covariate effect. Since no alternative models need to be specified, the residual-based test is applicable in many situations including LMMs and GLMMs. However, it may be less powerful than the other testing procedures that are specifically designed for testing a particular covariate effect. Comparison of the residual-based tests with the score tests in mixed models could be of future interest.

## Acknowledgements

The research of Daowen Zhang is partly supported by an NIH grant R01 CA85848-08. I would like to thank the referee and the managing editor Wendy Martinez for many valuable suggestions that greatly improved the presentation of this paper.